# Hydrogen gas sensing using aluminum doped ZnO metasurface

Sharmistha Chatterjee,[†,‡] Evgeniy Shkondin,[¶] Osamu Takayama,[§] Adam Fisher,[‡] Arwa Fraiwan,[∥] Umut A. Gurkan,[∥,⊥,#,@] Andrei V. Lavrinenko,[§] and Giuseppe Strangi∗,[‡,†]

[†]CNR-NANOTEC Istituto di Nanotecnologia and Department of Physics, University of Calabria, 87036 Rende, Italy.

[‡]Department of Physics, Case Western Reserve University, 10600 Euclid Avenue, Cleveland, OH 44106, USA.

[¶]DTU Nanolab - National Center for Micro- and Nanofabrication, Technical University of Denmark, Ørsteds Plads 347, DK-2800 Kgs. Lyngby, Denmark.

[§]DTU Fotonik – Department of Photonics Engineering, Technical University of Denmark, Ørsteds Plads 343, DK-2800 Kgs. Lyngby, Denmark.

[∥]Case Biomanufacturing and Microfabrication Laboratory, Mechanical and Aerospace Engineering Department, Case Western Reserve University, Cleveland, Ohio 44106, USA.

[⊥]Biomedical Engineering Department, Case Western Reserve University, Cleveland, Ohio 44106, USA.

[#]Department of Orthopedics, Case Western Reserve University, Cleveland, Ohio 44106, USA.

[@]Advanced Platform Technology Center, Louis Stokes Cleveland Veterans Affairs Medical Center, Cleveland, Ohio 44106, USA

E-mail: gxs284@case.edu

Phone: (216) 368 6918

**Abstract:** Hydrogen ($H_2$) sensing is crucial in a wide variety of areas, such as, industrial, environmental, energy and biomedical applications. However, engineering a practical, reliable, fast, sensitive and cost-effective hydrogen sensor is a persistent challenge. Here we demonstrate hydrogen sensing using aluminum-doped zinc oxide (AZO) metasurfaces based on optical read-out. The proposed sensing system consists of highly ordered AZO nanotubes (hollow pillars) standing on a $SiO_2$ layer deposited on a Si wafer. Upon exposure to hydrogen gas, the AZO nanotubes system shows a wavelength shift in the minimum reflectance by ~ 13 nm within 10 minutes for a hydrogen concentration of 4 %. These AZO nanotubes can also sense the presence of a low concentration (0.7 %) of hydrogen gas within 10 minutes. Its rapid response time even for low concentration, possibility of large sensing area fabrication with good precision, and high sensitivity at room temperature make these highly ordered nanotubes structures a promising miniaturized $H_2$ gas sensor.

**Introduction:** Hydrogen ($H_2$) gas is a highly combustible diatomic gas with low ignition energy. Therefore, when hydrogen gas leaks into external air, it may spontaneously ignite the environment and can cause an explosion, which is extremely hazardous. Hydrogen gas, within its concentration range of 4 to 75 % is considered combustible. [1] As $H_2$ is a tasteless non-toxic gas, it is hard to detect by human sense. Apart from the danger of explosion, hydrogen possesses a number of threats, such as asphyxiation in its pure, oxygen-free form, [2] frostbite ability associated with very cold liquid cryogenic hydrogen. [3] Therefore, for safety considerations, the detection of low hydrogen concentration within a very short time is crucial in industries where it is considered an obvious component or a byproduct.

Chromatography and mass spectrometry are widely used for large scale industrial $H_2$ gas sensing. However, these techniques are not suitable for use in miniaturized systems such as in the food industries and medical applications where the detection of ultra-low concentration of exhaling hydrogen has to be performed over a smaller area. [4-6] Until now, many groups reported some nanophotonic and nanoplasmonic system-based optical hydrogen sensors which work by controlling light at the sub-wavelength scale using their metallic components. [5, 7-8] However, in most cases, the metallic systems are characterized by high optical losses, while metal hydrides are of low loss but relatively more selective. Moreover, the sensing systems fabricated by lithography have low throughput, and inclined to imperfections and thus may be imperfect for optical detection of hydrogen gas. [9-10]

As losses are inherent to the plasmonic substance, alternative materials with lower optical losses result in being more suitable for sensing platforms. Zinc oxide (ZnO)-based nanostructures, [11] as a dielectric material, have been studied extensively for highly sensitive, selective and efficient gas sensors for the detection of various hazardous and toxic gases such as NO, $NO_2$,[12-13] $H_2$,[14-15] ammonia ($NH_3$),[16] methane ($CH_4$),[16-17] acetone,[18] ethanol,[19-22] humidity,[23] CO,[16, 24-26] volatile organic compound (VOCs),[27] and hydrogen.[17, 28-38] According to the literature, ZnO holds the promise to develop technologies based on resistive-type gas sensor for electrical readout. However, the high operating temperature, slow response time, poor selectivity and stability limit its extensive applications in the field of dissolved gas monitoring. For this electrical detection method the presence of $H_2$ gas is detected by monitoring the change of the resistivity of ZnO structures. On the contrary, optical detection is desirable not only because of the easy readout but also for the stringent safety conditions.

Here we experimentally demonstrate hydrogen sensing using metasurfaces made of aluminum-doped zinc oxide (AZO) nanotubes (hollow pillars). We have conducted a comparative study on the gas sensing property of the solid AZO pillars. The high aspect ratio AZO pillars and nanotubes were fabricated using a combination of advanced reactive ion etching and atomic layer deposition (ALD) techniques.[39] The solid AZO pillars do not respond to the $H_2$ gas of concentration 0.7 % - 4 % at room temperature and pressure even after a long exposure time. On the other hand, the AZO nanotubes show a wavelength shift of ~ 13 nm upon exposure to $H_2$ gas of 4 % concentration at ambient temperature and a simultaneous decrease in the reflectance intensity of ~ 0.4 % is also detected within 10 minutes. This structure can even sense the presence of low concentration (0.7 %) of $H_2$ gas within the same response time. In order to understand the sensing mechanism of the AZO metasurfaces, numerical simulations have been performed to simulate the experimental behavior of the AZO nanotubes exposed to $H_2$ gas. In particular, the following paragraphs describe the design and the nanofabrication process of the metasurfaces along with the finite element method (FEM) based Comsol simulations to understand the peculiar aspects of the interaction of light waves onto the metasurfaces. Then the sensing experiments are described starting by the engineering of microfluidic gas flow chambers.

**EXPERIMENTAL SECTION**

**Fabrication of AZO Nanotubes:** Two types of samples were fabricated: AZO solid pillars and nanotubes structures on Si substrate with thermally grown 200 nm silica layer (see Figure 1). The external diameter, thickness, and height of the AZO nanotubes are 300 nm, 20 nm, and 2 μm respectively and a pitch of 400 nm is maintained over 1×1 $cm^2$ area. On the other hand, the diameter and height of the AZO solid pillars are 300 nm and 2 μm, respectively, and a pitch of 400 nm is also maintained here over the same area. For both types of structures, air acts as the host material. Elemental analysis by transmission electron microscopy (TEM) and other characterization methods as well as extended details on fabrication of AZO pillars and nanotubes can be found elsewhere.[39]

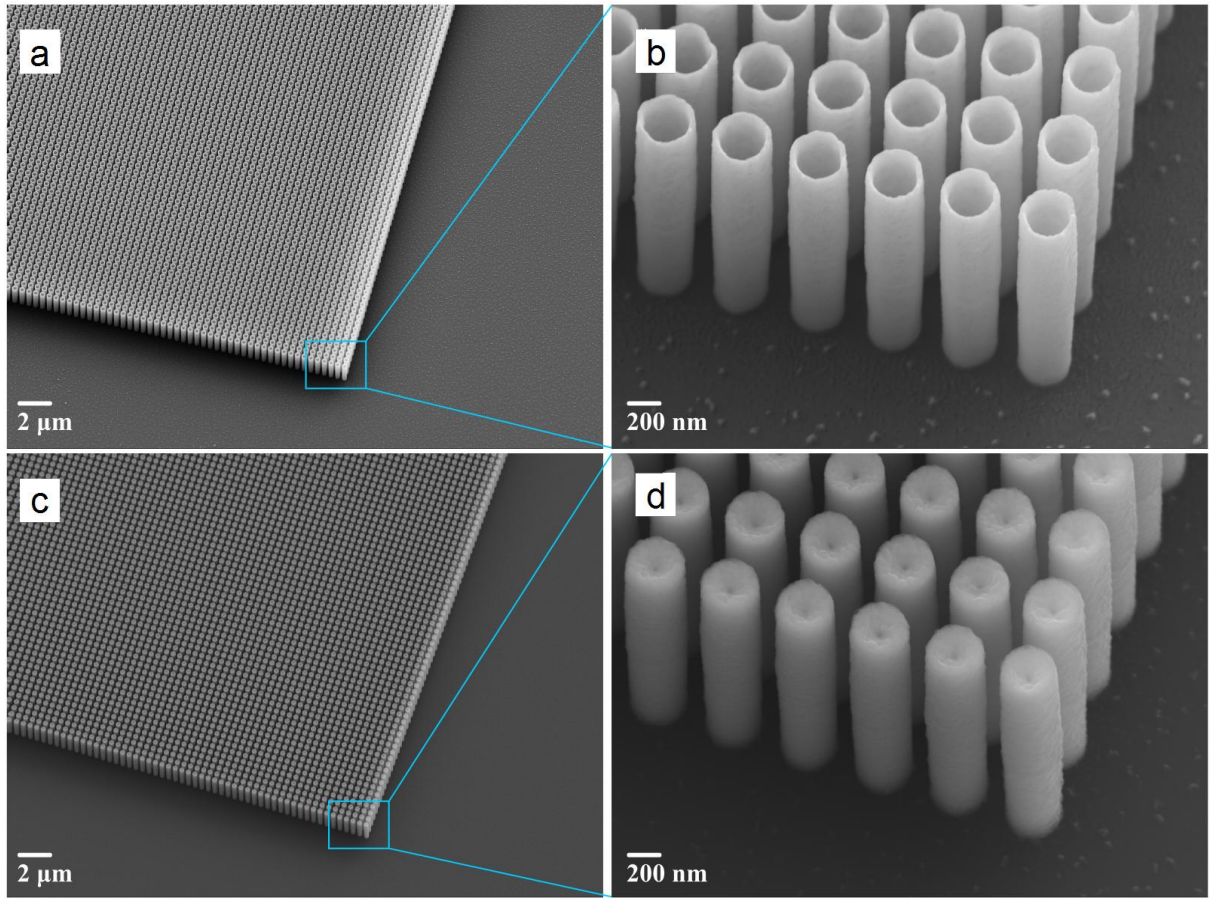

Figure 1: Scanning electron microscope (SEM) images of fabricated AZO (a, b) nanotubes and (c, d) pillar structures with a pitch of 400 nm, diameter of 300 nm, and height of 2 μm. The wall thickness of nanotubes is approximately 20 nm.

Deep-UV lithography was used to define grating patterns on standard silicon <100> wafers. DRIE was implemented with a standard Bosch process [40] in order to fabricate a Si template with the 2 μm-deep air hole arrays. Afterwards, the processed structures were cleaned in $N_2/O_2$ plasma in order to remove resist remaining and other organic contaminants. Then, the silicon templates were coated by AZO (using trimethylaluminum, diethylzinc, and water as precursors) by means of atomic layer deposition (ALD), until the air holes were filled entirely. ALD is based on the self-limiting, sequential surface chemical reactions which allow conformal deposition on complex structures with thickness control. [41] For the final step, the AZO filling needs to be isolated, and for that purpose, the samples were subjected to $Ar^+$ sputtering for removal of the top AZO layer and exposing the silicon template. Afterwards, the silicon template in-between cavities coated with AZO, has been removed using $SF_6$ plasma in conventional isotropic reactive ion etching process without interference with functional ALD material, resulting in the formation of AZO pillars and tubes as shown in Figure 1. The more detailed description of the fabrication method for different structures, such as, AZO trenches, [42-44] TiN trenches, [45] dielectric trenches, [46] coaxial tubes [47] can be found elsewhere. The optical properties of

AZO films fabricated by ALD have been measured by spectroscopic ellipsometer for the wavelength range of interests. [39, 42]

**Finite Element Method (FEM) Simulation:** Finite Element Method (FEM) simulation is carried out to describe the interaction between light and AZO nanotubes with 400 nm periodicity in square lattice arrangement. In order to map out the electric field pattern in the nanotubes structures interacting with light we used the Radio Frequency (RF) module of FEM based COMSOL 5.4 software package. During this theoretical study, all the specifications about the size, shape, substrates of AZO nanotubes are obtained from the SEM results. Moreover, the COMSOL material library is used for the optical properties of air as surrounding medium, $SiO_2$ layer, and Si substrate. The wavelength dependent complex permittivity with both real and imaginary parts of AZO is taken from the experimental data obtained from spectroscopic ellipsometry measurements and a proper fitting. [39] Linearly polarized plane waves are used as the excitation source. The direction of the electric field of the incident light is perpendicular to the semi major axis of the nanotubes (TE-polarized light) which is identical to the experimental framework. Physics controlled free triangular and tetrahedral meshes (depending on different layers) are used for these analysis to calculate the electromagnetic power loss density ($W/m^3$) and the electric field enhancement with a suitable and variable mesh size where the lowest mesh size is kept at 1.6 nm.

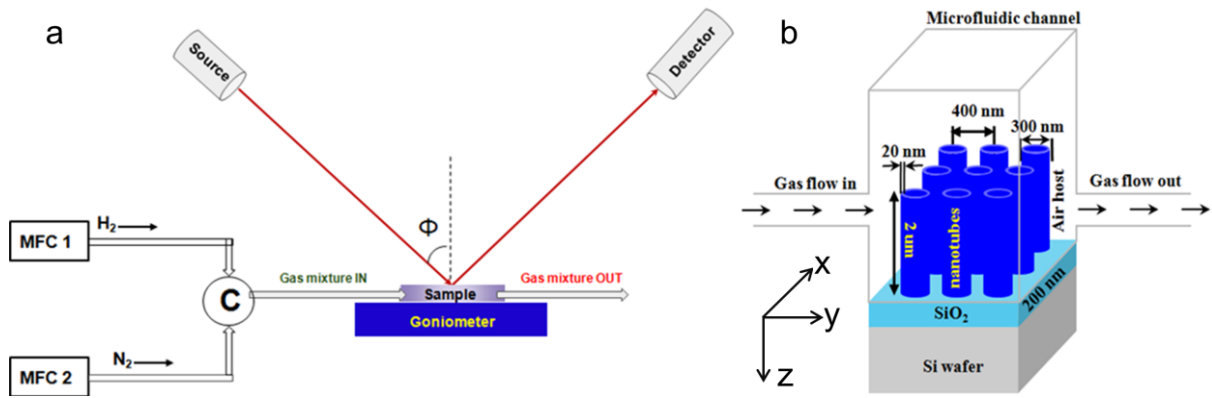

Figure 2: (a) Schematic diagram of the experimental set-up. (b) Schematic diagram of the AZO nanotubes sensing system.

**Fabrication of Microfluidic Channel:** The microfluidic channel is made up of poly-(methyl methacrylate) (PMMA) plastic window which is surrounded by the micromachined inlets and outlets. Along with that a double-sided adhesive film is attached there which defines the outlines and thickness of the microfluidic channels. The diameter of inlets and outlets of the PMMA top is 0.61 mm and separation distance of 12.4 mm is maintained here. The PMMA top is fabricated by laser micromachining using a VersaLASER system (Universal Laser Systems). A double-sided adhesive film (iTapestore, 100 μm in height) has been laser micro-machined to have the same precise size of PMMA top and is placed within the 14×2 mm micro-channel. Here the adhesive film is attached to the PMMA top to include the inlet and outlet within the outline of the channel. To connect the gas sources with the microfluidic channel fluorinated ethylene propylene (FEP) tubing (Cole-Parmer) is used. All the tubings and their connections between gas sources and channels are fastened using a 5 min epoxy (Devcon). For clarity a picture of our sensor (Figure S1) is given in the Supporting Information.

**Ellipsometric Reflection Measurement for Gas Sensing:** A high-resolution variable angle spectroscopic ellipsometry (SE) (J. A. Woollam Co., Inc., V-VASE) is used to experimentally measure all types of angular reflection data as illustrated in Figure 2a. The low-power spectroscopic ellipsometer has high precision and it is non-destructive and thus preferable for the optical characterization of nanostructures and gas sensing. In order to characterize the performance of our gas sensing system, AZO nanotubes are exposed to $H_2$ gas of different concentrations. The channel inlet is connected to a tank of nitrogen mixed with 4 % volume concentration of $H_2$ and with pure nitrogen. The gas flow from both the tanks is controlled separately using two separate mass flow controllers (MFCs). By adjusting those MFCs, we could control the hydrogen concentration (between 0 and 4 %) introduced to the sample as illustrated in Figure 2b. After adjusting all these parameters the sensor was placed onto the variable angle ellipsometer to measure the reflectivity $R(\lambda, \phi)$ at different angle of incidence, $\phi$, in the wavelength range of $\lambda = 300 - 1500$ nm. The maximum resolution of this ellipsometer is 0.03 nm but for our measurements 1 nm resolution is maintained throughout the experiments. Note that, in this work we present experimental results of TE-polarized incident light because TM-polarized light does not show any sensitivity toward hydrogen gas (see Figure S2 in Supporting Information). During gas sensing experiment, because of the PMMA window on top of the channel, the incident beam splits into two reflected beams at the interfaces of Air-PMMA window and PMMA window-micro fluidic channel, causing Fabry-Perot fringes in the reflection spectra. In order to minimize the effect and keep the associated error minimum, a particular suitable incident angle for the sensing set-up needs to be optimized. Here the measurement angle is kept below or equal to $\phi = 45^o$. Moreover, the iris of the ellipsometer detector helps to block the additional reflected beam during the sensing measurement, an unavoidable source of experimental error which originates from the insertion of the micro fluidic channel. After fixing the measurement angle to see the response of the AZO sensing system for $H_2$ gas of a particular concentration, the reflectance has been continuously measured over a wavelength range (selected based on the minima of reflectance) for 90 minutes keeping a constant time interval between each measurement.

## RESULTS AND DISCUSSION

**Finite Element Method (FEM) Simulation:** Figure 3a shows the simulated reflection from AZO nanotubes. The incident angle is taken as $\phi = 45^o$ similar to the experimental case. Here 400 nm periodicity, 20 nm thick tube wall, and 2 μm height are considered. The direction of the electric field of the incident light is taken along x-axis (TE-polarized incident light). The inset in Figure 3a shows the numerically calculated shift of the 1100 nm mode in reflectance spectra after $H_2$ gas injection.

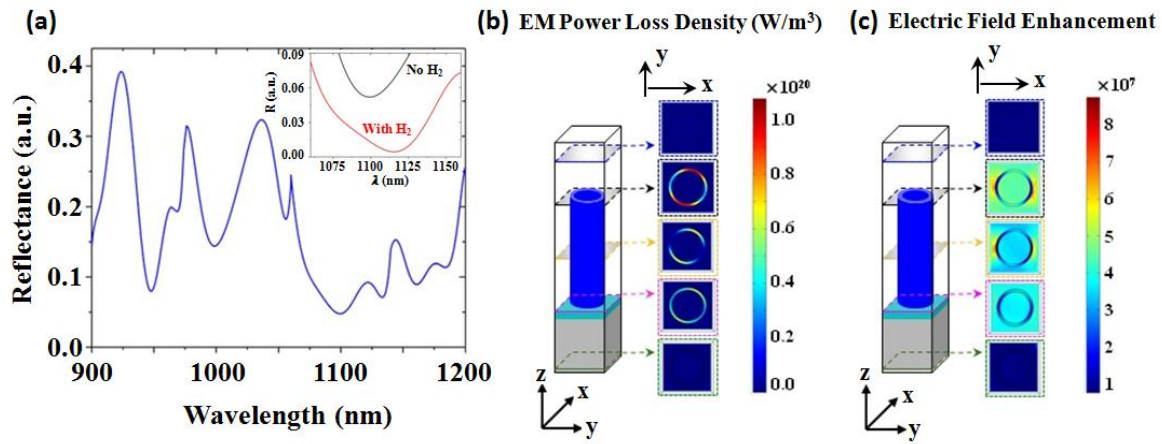

Figure 3: Simulation results. (a) Reflection spectra of AZO nanotubes array system. (b) The electromagnetic power loss density (W/m$^3$) and (c) Electric field prole at 1100 nm in wavelength. The inset figure shown in panel (a) depicts the change in the 1100 nm mode of reflectance spectra after intercalation of $H_2$ gas. Here simulation for linearly polarized incident plane wave (TE-polarized light with electric field parallel to the surface of tubes along x-axis) is investigated. The periodicity, thickness of wall, and height of tubes are 400 nm, 20 nm, and 2 µm, respectively. The incident angle is $\phi = 45$.

During simulation the refractive index of the hydrogen is taken as 1.00014.[48] The wavelength shift in this case is 17 nm and the change in intensity of the reflectance minima mode is 0.048 (see Figure S3 in Supporting Information for more details). The electromagnetic power loss density (W/m$^3$) associated with the AZO nanotubes for plane incident waves are shown in Figure 3b at 1100 nm, which is one of the reflection minima modes in the theoretical reflectance spectra of AZO nanotubes shown in Figure 3a. Here, it is clear that the optical loss is concentrated mostly within the AZO nanotubes. The light absorption in the AZO tube layer suggests that the AZO tubes absorption properties can be significantly modified by the environmental conditions. Simulated electric field profile at the wavelength of 1100 nm is shown in Figure 3c which depicts the high electric field enhancement near the surface of AZO nanotubes, useful for the detection of any changes of AZO nanotubes optical properties.

**Hydrogen Sensing by AZO Nanotubes:** We study the spectral shift, $\Delta\lambda$, for different hydrogen concentrations. Figure 4a shows the response of AZO nanotubes system for 0.7 % $H_2$ gas at $\phi = 45°$ over a wavelength range of $\lambda = 300$ nm - 1500 nm. This plot shows that for a particular measurement angle the wavelength shift, $\Delta\lambda$, is larger for longer wavelengths, giving higher sensitivity. For example, for $\phi = 45°$, the wavelength shift of reflectance minima around $\lambda = 900$ nm is $\Delta\lambda = 0$ nm, while reflectance minima at $\lambda = 1126$ nm shifts by $\Delta\lambda = 2.6$ nm for 0.7 % $H_2$ gas. Depending on the angle of incidence for measurement, the sensitivity can also be tuned. More detailed information about the dependence of wavelength shift and sensitivity on the measurement angle variation is given in Supporting Information (Figure S4). We note that for longer wavelengths the fluctuation in reflectance increases due to Fabry-Perot interference originating from the PMMA window of micro fluidic channel. Taking these factors into account, here we focus on the reflection minima around 1100 nm in wavelengths for hydrogen sensing.

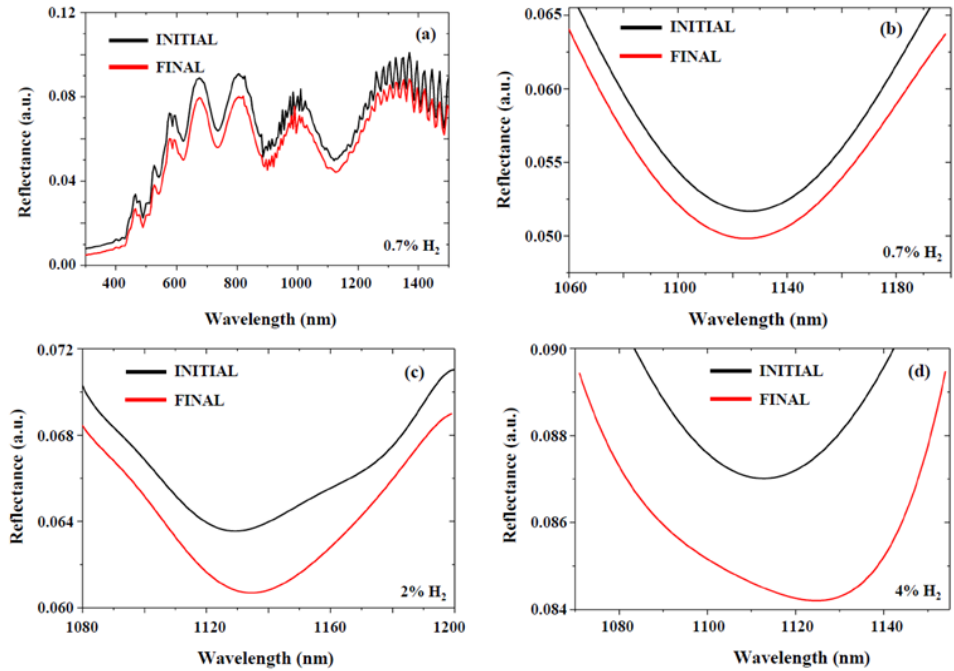

Figure 4: The response of AZO nanotubes in presence of hydrogen gas. (a) The response of AZO nanotubes sensing system before and after intercalation of 0.7 % $H_2$ gas measured at the incident angle, $\phi = 45°$ over a wavelength range of $\lambda = 300$ nm - 1500 nm. Using this plot one can choose the suitable mode for $H_2$ gas detection with higher sensitivity. (b) 0.7 %, (c) 2 %, and (d) 4 % $H_2$ gas sensing results. For all these measurements, the response time is 10 min.

In Figure 4b - 4d, properly fitted $H_2$ gas sensing response of AZO nanotubes array system is presented for concentration 0.7 %, 2 %, and 4 %, respectively. For all these measurements the response time was 10 minutes. Here, the system shows wavelength shift, $\Delta\lambda$, and reflectance intensity change, $\Delta I$, even for the lowest concentration 0.7 % of $H_2$ gas. The wavelength shift is 2.6 nm for 0.7 % $H_2$ and the change in reflectance intensity is 0.2 %. Similarly, the wavelength shift is 5 nm for 2 % $H_2$ and the change in reflectance intensity is nearly 0.3 %. For 4 % $H_2$ gas the wavelength shift is 13 nm and the reflectance intensity change is 0.4 %. In comparison to the theoretically calculated wavelength shift (17 nm) and the intensity change (0.0475) of the reflectance minima mode, the experimental changes even for the case of 4% $H_2$ gas, are little bit smaller. This is because of the consideration of ideal situations in case of theoretical analysis.

When light activated-AZO nanotubes interact with the atmospheric air, oxygen molecules get absorbed on the surface of the nanotubes which further extract electrons from the conduction band of the ZnO site. When a reducing gas like $H_2$ intercalates the system, those light-activated chemisorbed oxygen ions on the surface of AZO nanotubes interact with the reductive gas molecules and donate free electrons back to the conduction band of ZnO and produce water molecules.[49] This local change in the polarizability and thus the surrounding refractive index contributes to the shift of reflection minima in presence of $H_2$ gas. Here it is worth mentioning that the metal doping in ZnO nanotubes increases the photo-generated free electron-hole pairs and because of the resonant plasmonic effect, light absorption capacity is enhanced which further results in more sites for the oxygen molecules on the surface of AZO nanotubes.[49] The AZO nanotubes-based metasurface as a Fabry-Perot cavity senses the presence and concentration of $H_2$ gas by detecting the change in polarizability due to the chemical changes in the surrounding media.

If wavelength shift is considered as the measurement parameter, the signal to noise ratio (S/N) becomes 16.7 % which is equivalent to the wavelength shift / FWHM (Full Width at Half Maxima) of the mode. In order to characterize the sensitivity in terms of reflectance change, the figure of merit (FOM) is defined as FOM = R ($AZO_{air}$)/R ($AZO_{H2}$) ×100,[8] where R ($AZO_{air}$) is the reflectance of the sensor in $N_2$ (which is equivalent to air media) without hydrogen and R ($AZO_{H2}$) is the reflectance of the sensor in presence of hydrogen gas. According to Figure 4d for 4 % $H_2$ gas, the calculated FOM of our sensor is FOM ~ 104. Note that we also conducted reflectance measurement for AZO solid pillar in presence of 4 % $H_2$ gas over 90 minutes. However, even after such a long time, there is no clear change observed in the reflectance minima mode in terms of wavelength shift or reflectance intensity change for such gas sensing with the solid AZO pillars (see Figure S5 in Supporting Information). This indicates that the ultra thin wall thickness of AZO nanotubes and their hollowness play an important role in sensing for its larger surface area to volume ratio relative to the solid pillars.[50] Here it is worth mentioning that, for AZO nanotubes (hollow pillars), the surface area to volume ratio is nearly 13.2 times larger than that of the solid AZO pillars.

Figure 5a and 5b display the wavelength shift and the intensity change in the reflectance minima of the AZO metasurface after exposure to $H_2$ gas with different concentration. Based on the measurement of wavelength shift, here the limit of detection (LOD) is 0.2 % $H_2$ for 10 minutes of response time considering the fact that the maximum possible resolution of the used spectroscopic ellipsometer is 0.03 nm (See Figure S6 in Supporting Information). It is well-known that practical hydrogen sensors must be able to respond to hydrogen at concentrations approximately an order of magnitude lower than the explosive limit of $H_2$ (4%).[8, 49] The previous work shows Pd-based optical $H_2$ gas sensor with high FOM in room temperature.[8] However, the response time was nearly 60 minutes, which can be an issue for the safety in the industrial field. Our AZO nanotubes array system could achieve the requirement in terms of fast response time of 10 minutes at room temperature by means of safe optical measurement.

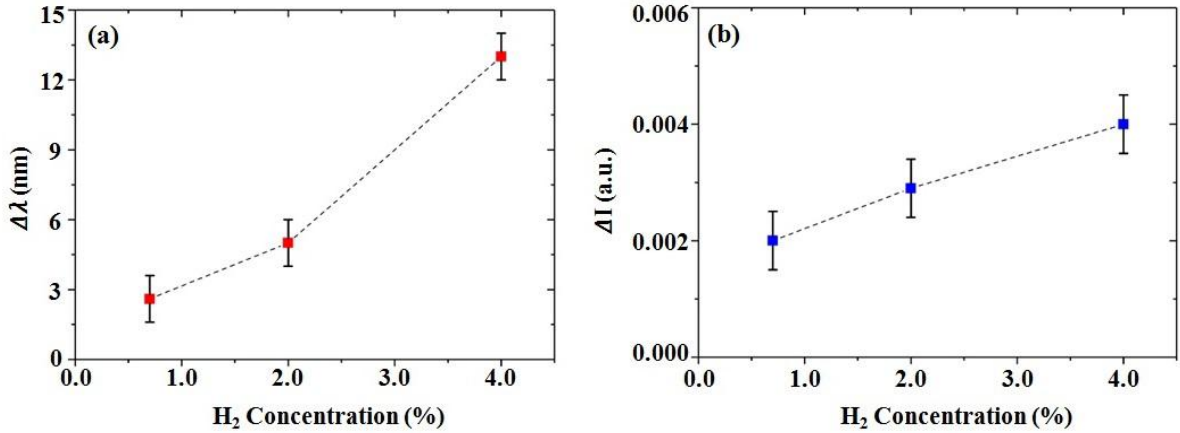

Figure 5: Variation of experimental (a) wavelength shift, $\Delta\lambda$, and (b) reflectance intensity change of reflection minima, $\Delta I$, observed in AZO nanotubes sensing system in presence of $H_2$ gas of different concentration (0.7 %, 2 %, and 4 %). Here the dotted lines are for the guidance to see how $\Delta\lambda$ and $\Delta I$ changes for different concentration of $H_2$ gas.

The results of gas sensing response over time up to 90 minutes for 4 % $H_2$ gas is shown in Figure 6a from which we can see the fast response nature of the AZO nanotubes sensor. Here it can be seen that 10 minutes is sufficient to detect 4 % $H_2$ gas. This response time is also valid for 2 % and 0.7 % of $H_2$ gas. Figure 6b shows the response of AZO nanotubes in different conditions. Here the ellipsometric response of the AZO nanotubes system with PMMA channel in absence of any gas as well as in presence of pure $N_2$ gas and 4 % $H_2$ gas are shown.

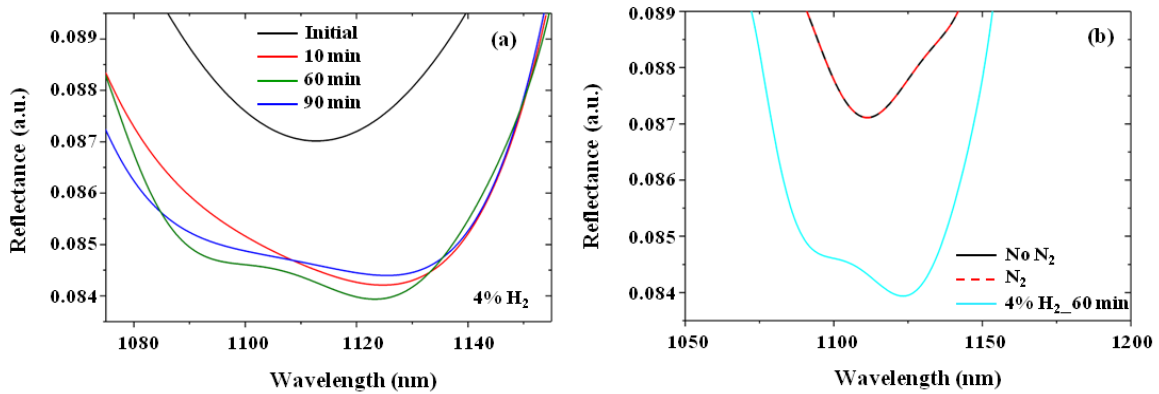

Figure 6: (a) Response of AZO nanotubes sensing system overtime in presence of 4 % $H_2$ gas. It is very clear that 10 min is enough to detect the presence of 4 % $H_2$ gas. (b) The response of AZO nanotubes sensing system (with PMMA channel) in presence and absence of $N_2$ gas where there is almost no variation in the mode. But a clear shift of the mode is observed in presence of 4 % $H_2$ gas after 60 min.

In this plot, we do not observe any wavelength shift in reflection after introducing pure $N_2$ gas to the sensor (dashed red curve) compared to the reflectance when the sample is not exposed to any gas (black curve). In comparison, after introducing $N_2$ mixed with 4 % of $H_2$ (cyan curve), we observe a clear red shift in the reflection minimum, as well as a change in the reflection amplitude due to the intercalation of hydrogen atoms in the sensing system as expected. By introducing pure $N_2$ to the channel again, the reflectance did not go back to its primary position. However, the absorption properties of the AZO nanotubes system remain same for a longer time afterward. Here different cycles of introducing $N_2$ mixed with 4 % of $H_2$, and pure $N_2$, show almost no change in the spectral properties of the absorption mode of AZO nanotubes sensor.

**CONCLUSION**

In summary, we report the study of $H_2$ gas sensing properties of a metasurface made up of highly ordered high aspect ratio aluminium-doped zinc oxide (AZO) nanotubes designed to operate in the NIR range, around 1100 nm. While AZO solid pillars do not respond to $H_2$ gas, AZO nanotubes show a wavelength shift as well as a significant decrease in the reflectance intensity after exposing them to $H_2$ gas. Here $H_2$ gas sensing is observed at room temperature and pressure within a response time of 10 minutes. Therefore, the possibility of having a room-temperature $H_2$ gas sensor, with all the advantages of low-response time, capability to detect low concentration of $H_2$ gas, a good figure of merit, possibility of fabricating a large area sensor with high precision and reliability can make these metasurfaces the platform for very promising $H_2$ gas sensors for industrial applications. Moreover, the sensitivity of nanotubes may be further enhanced by controlling the doping level of ZnO by aluminum. AZO metasurfaces hold the promise of realizing room temperature gas sensor, which may enable monitoring hydrogen gas by color changes in the visible range.

**AUTHOR INFORMATION**

**Corresponding Author**

*E-mail: gxs284@case.edu

**Note**

The authors declare no competing financial interests.

**Author Contributions**

S. C. wrote the manuscript, conducted modeling and theoretical analysis of the structures. S. C. and A. F. Performed optical characterization for hydrogen sensing. E. S. fabricated the samples. A. F. and U. A. G. built the microfluidic cell. O. T. and A. V. L. contributed to the discussions of results and manuscript writing. G. S. conceived the idea and supervised the work. All authors have given approval to the final version of the manuscript.

**Acknowledgement**

This work was supported by Independent Research Fund Denmark, DFF Research Project 2 "PhotoHub" (8022-00387B), Villum Fonden "DarkSILD project" (11116), Direktor Ib Henriksens Fond, Denmark, and NSF Grant (Award Number 1904592) - Instrument Development: Multiplex

Sensory Interfaces Between Photonic Nanostructures and Thin Film Ionic Liquids. The authors would like to acknowledge the support from the National Centre for Nano Fabrication and Characterization (DTU Nanolab).